%% file: main.tex
\documentclass[lettersize,journal]{IEEEtran}
\usepackage{amsmath,amsfonts}
\usepackage{algorithmic}
\usepackage{algorithm}
\usepackage{array}
\usepackage[caption=false,font=normalsize,labelfont=sf,textfont=sf]{subfig}
\usepackage{textcomp}
\usepackage{stfloats}
\usepackage{url}
\usepackage{verbatim}
\usepackage{graphicx}
\usepackage[switch]{lineno}
\usepackage[dvipsnames]{xcolor}
\usepackage[normalem]{ulem}
\usepackage{booktabs}
\usepackage{siunitx}

\begin{document}
% \linenumbers

\title{A Multi-Channel Auditory Signal Encoder with Adaptive Resolution Using Volatile Memristors}

\author{Dongxu~Guo,
Deepika~Yadav,
Patrick~Foster,
Spyros~Stathopoulos,
Mingyi~Chen,
Themis~Prodromakis,
and Shiwei~Wang%

\thanks{D. Guo, D. Yadav, P. Foster, S. Stathopoulos, T. Prodromakis, and S. Wang are with the Institute for Integrated Micro and Nano Systems (IMNS), School of Engineering, The University of Edinburgh, Edinburgh EH9 3FF, U.K. (e-mail: dongxu.guo@ed.ac.uk).}%
\thanks{M. Chen is with the Department of Micro/Nano Electronics, Shanghai Jiao Tong University, Shanghai 200240, China.}%}
}

\markboth{This work has been submitted to the IEEE TCAS1 for possible publication.}%
{}


\maketitle

\begin{abstract}

\input{contents/abstract.tex}

\end{abstract}

\begin{IEEEkeywords}
Neuromorphic Systems, Adaptive Threshold, Volatile Memristors, Asynchronous Delta Modulator (ADM), Short-Term Adaptation, Spike Encoding
\end{IEEEkeywords}

\input{contents/intro.tex}

\input{contents/short_term_adaptation.tex}

\input{contents/system_architecture.tex}

\input{contents/implementation.tex}

\input{contents/measurement_results.tex}

\input{contents/conclusion.tex}

% \input{contents/Fabrication.tex}

% \input{contents/appendix.tex}


\newpage

\bibliographystyle{IEEEtraN}
\bibliography{mybib}

\newpage

 
\vspace{11pt}

\vfill

\end{document}

%% file: contents/abstract.tex
We demonstrate and experimentally validate an end-to-end hybrid CMOS–memristor auditory encoder that realises adaptive-threshold, asynchronous delta-modulation (ADM)-based spike encoding by exploiting the inherent volatility of HfTiO$_x$ devices. A spike-triggered programming pulse rapidly raises the ADM threshold $\Delta$ (desensitisation); the device’s volatility then passively lowers $\Delta$ when activity subsides (resensitisation), emphasising onsets while restoring sensitivity without static control energy. Our prototype couples an 8-channel 130 nm encoder IC to off-chip HfTiO$_x$ devices via a switch interface and an off-chip controller that monitors spike activity and issues programming events. An on-chip current-mirror transimpedance amplifier (TIA) converts device current into symmetric thresholds, enabling both sensitive and conservative encoding regimes. 
Evaluated with gammatone-filtered speech, the adaptive loop—at matched spike budgets—sharpens onsets and preserves fine temporal detail that a fixed-$\Delta$ baseline misses; multi-channel spike cochleagrams show the same trend. Together, these results establish a practical hybrid CMOS–memristor pathway to onset-salient, spike-efficient neuromorphic audio front-ends and motivate low-power single-chip integration.

%% file: contents/intro.tex
\section{Introduction}
\IEEEPARstart{M}{imicking} the structure of biological neural networks has been a driving force behind the development of artificial neural networks (ANNs). 
ANNs have been scaling to increasingly larger sizes on digital hardware, leading to significant improvements in various applications such as speech recognition, 
natural language processing, and autonomous systems. However, the growing complexity and size of these models have raised significant concerns regarding their energy efficiency.

Biological neural networks perform complex computations with minimal energy consumption by encoding information into sparse spikes and using these spikes for communication and processing.
This observation has inspired researchers to explore more energy-efficient computing paradigms, such as spiking neural networks (SNNs), 
which aim to replicate the event-driven nature of biological neurons.

A fundamental aspect of SNNs and neuromorphic systems is the process of spike encoding, 
where analogue sensory signals are transformed into sequences of discrete spikes. This spike-based representation enables efficient, 
event-driven processing and is central to the remarkable energy efficiency observed in biological systems. 
Implementing effective spike encoding in hardware is thus a critical step towards realising practical neuromorphic processing systems.

One of the main challenges in spike encoding is minimising the number of spikes to save processing energy while preserving as much information as possible. 
For auditory signal encoding, asynchronous delta modulators (ADMs) have proven effective in many engineered systems\cite{yang20160}\cite{ACM_delta}\cite{koickal2011design}\cite{guo2020asynchronous}, 
as they resemble the phase-locking property of auditory nerve fibres\cite{dreyer2006phase} and operate in an event-based manner. To preserve more information, a delta modulator must have a 
smaller $\Delta$, ensuring sufficient resolution\cite{sayiner2002level}. However, this inevitably causes larger spike counts.

Various works from different communities have tried to address this problem. Considering the delta modulator as a Level-Crossing ADC (LCADC), the total spike count can be reduced by temporarily increasing $\Delta$ when a level-crossing is triggered\cite{weltin2013event}\cite{alea2024fingertip}. From a system perspective, a neuromorphic system could modulate
the spike count by adjusting the gain of the amplification stage preceding the delta modulator\cite{yang20160}\cite{kiselev2022spiking}, mimicking the adaptation found in ossicles and outer-hair cells.
While these approaches have proven effective in one way or another, an intrinsic property of auditory nerve fibres—short-term adaptation—has not been explored to solve this encoding problem.

Short-term adaptation is a well-observed phenomenon in auditory nerve fibres\cite{perez2014adaptation}: 
in response to a sustained sound stimulus, the firing rate initially peaks and then gradually decreases, 
with time constants on the order of tens of milliseconds. 
This local automatic gain control (AGC) mechanism has been shown to enhance onset responses and is essential for accurate speech recognition\cite{koning2016speech}.

Utilising this concept in delta modulation, the encoding scheme operates as follows: 
the system initially sets a very small threshold ($\Delta$), resulting in a maximum response. 
When a sound occurs, the threshold is dynamically increased, leading to a controlled reduction in the response. 
After a brief period, the threshold gradually returns to its minimum value spontaneously, preparing the system for the next sound burst. 
This adaptive mechanism enhances the detection of speech onsets while reducing the overall spike count, thereby improving spike-conversion efficiency.

Memristors provide an efficient hardware mechanism for adaptive thresholding and short-term memory, making them well suited for speech onset enhancement in delta modulation. A memristor is a two-terminal device characterised by a pinched hysteresis loop in the current–voltage ($I$–$V$) plane, reflecting its resistive memory behaviour\cite{1083337}. Depending on retention characteristics, memristors are broadly classified as non-volatile, which preserve resistive states after bias removal, and volatile, which gradually relax back to a high-resistance state. While non-volatile devices are attractive for storage and in-memory computing, volatile devices offer the intrinsic decay essential for transient memory and adaptive processing\cite{wang_recent_2020}. This natural relaxation is analogous to sensory adaptation in biological systems, enabling volatile memristors to act as compact, low-power building blocks for short-term adaptation in neuromorphic circuits.

In this work, we present an 8-channel auditory signal encoder on silicon that incorporates a memristor-assisted short-term adaptation mechanism to enhance speech onset detection and improve spike encoding efficiency. 
By leveraging the intrinsic volatility of memristors, our design emulates the adaptive firing behaviour of biological auditory nerve fibres, providing a local and energy-efficient automatic gain control. 
This bio-inspired approach enables dynamic sensitivity adjustment to salient temporal features, such as sound onsets, 
which are critical for speech perception.

The rest of the paper is organised as follows: Section II provides an overview of short-term adaptation in auditory systems and the rationale for using volatile memristors to mimic it. 
Section III describes the architecture of the proposed encoding system. Section IV details the design and implementation of the system, and Section V presents experimental results demonstrating the effectiveness of our approach. 
Finally, Section VI concludes the paper and discusses future directions.

%% file: contents/short_term_adaptation.tex
\section{Short-Term Adaptation and Volatile Memristors}
Short-term adaptation is an intrinsic property of auditory nerve fibres' response, 
where the firing rate initially peaks in response to a sound stimulus and then gradually decreases over time.
This phenomenon has been shown to be essential for speech understanding, especially in adverse conditions, by incorporating this feature into the signal processing of conventional
cochlear implants\cite{koning2016speech}\cite{azadpour2016enhancing}.

It is believed that the high-pass property of this phenomenon enhances the onset of consonants, which are weak in strength and easy to miss, particularly in challenging
listening environments. By incorporating short-term adaptation into cochlear implants (CI), CI users tend to perform better in speech-related tasks\cite{azadpour2016enhancing}.
The s-domain model used in \cite{azadpour2016enhancing} to represent this phenomenon is shown in Eq.\ref{eq:stp} ($\rho$ is the onset enhancement gain, and $\tau$ is the decay time constant). 
The model demonstrates that short-term adaptation is essentially the initial input plus an enhancement of the fast-changing edge at onset.
\begin{align}
    \label{eq:stp}
    H(s) &= \rho \frac{\tau s + \rho^{-1}}{\tau s + 1} \notag \\
         &\approx  1 + \frac{\rho s}{\tau s + 1} \quad \text{(considering $\tau \ll \rho$)}
\end{align}

% Short-term adaptation in biology finds a close parallel in volatile memristors \cite{ricci_decision_2022}. Moreover, networks of volatile resistive memories have been shown to perform collective decision-making and onset-salient sensing, reinforcing their promise as neuromorphic front-ends for auditory encoding \cite{zhou2022volatileNonvolatileAELM}\cite{zuo2023volatileTS}\cite{brivio2022HfO2NCE}. Volatile memristors reduce their resistance under electrical stimulation and automatically recover to the high-resistance state once the input is removed, a behavior originating from the dissolution of nanoscale conductive pathways (e.g., Ag/Cu nanoclusters or sub-filaments) driven by thermodynamic and ionic relaxation \cite{wang2017diffusiveNatMat}\cite{zhou2022volatileNonvolatileAELM}\cite{zuo2023volatileTS}. Both the amplitude and duration of the input pulse strongly modulate the relaxation constant, with recovery times ranging from sub-$\mu$s to $\sim$s well aligned with biological adaptation timescales \cite{zuo2023volatileTS}\cite{chekol2022relaxAFM}. This relaxation provides a direct hardware analogue of the decay term in Eq.~\ref{eq:stp}, mapping volatile memristor dynamics onto auditory short-term adaptation. While HfOx based memristors can exhibit volatile behavior depending on electrode configuration, in our work we kept both the electrodes as TiN and instead engineered the dielectric by introducing Ti into the hafnia matrix, forming HfTiO$_x$ films that exhibit controlled volatility \cite{saylan_effects_2020}\cite{li2020HfO2selectorAdvSci}.

Volatile memristors provide a direct hardware analogue of short-term adaptation by harnessing an intrinsic material property—the volatility of the active oxide\cite{ricci_decision_2022}. Volatile memristors, which include threshold-switching selectors and diffusive devices, exhibit a stimulus-dependent conductance that relaxes back to a high-resistance state once excitation is removed. This volatility originates from thermodynamic/ionic relaxation of nanoscale conductive pathways (e.g., Ag/Cu nanoclusters or sub-filaments), enabling short-term, adaptive dynamics useful for neuromorphic front-ends \cite{wang2017diffusiveNatMat}\cite{zhou2022volatileNonvolatileAELM}\cite{zuo2023volatileTS}. In HfO$_2$-based stacks (and related mixed-oxide variants such as HfTiO$_x$), both the amplitude and width of a brief programming pulse set the subsequent decay: stronger excitation lengthens passive relaxation, with recovery constants spanning from sub-$\mu$s to $\sim$s across reported volatile devices \cite{zuo2023volatileTS}\cite{chekol2022relaxAFM}. Closely related HfO$_2$-selector structures show \emph{bipolar} threshold switching with spontaneous return (ns–$\mu$s spontaneous return reported) to the OFF state on device- and pulse-dependent timescales, underscoring the maturity and reproducibility of these mechanisms in hafnia-derived oxides \cite{li2020HfO2selectorAdvSci}. Comprehensive reviews place these dynamics within established materials/physics taxonomies and highlight their utility for onset-salient sensing and short-term synaptic emulation \cite{zhou2022volatileNonvolatileAELM}\cite{zuo2023volatileTS}\cite{brivio2022HfO2NCE}.

\begin{figure}[h]
\centering
\includegraphics[width=0.45\textwidth]{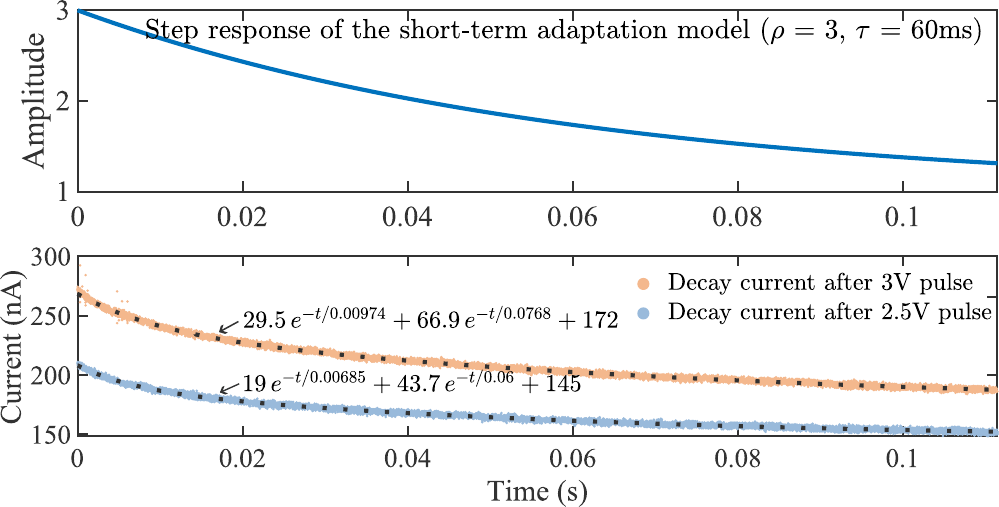}
\caption{(top) The step response of the short-term adaptation model; (bottom) Measured decay current of a volatile memristor with bi-exponential fit.}
\label{fig:STP and RRAM}
\end{figure}

\begin{figure}[h]
\centering
\includegraphics[width=0.45\textwidth]{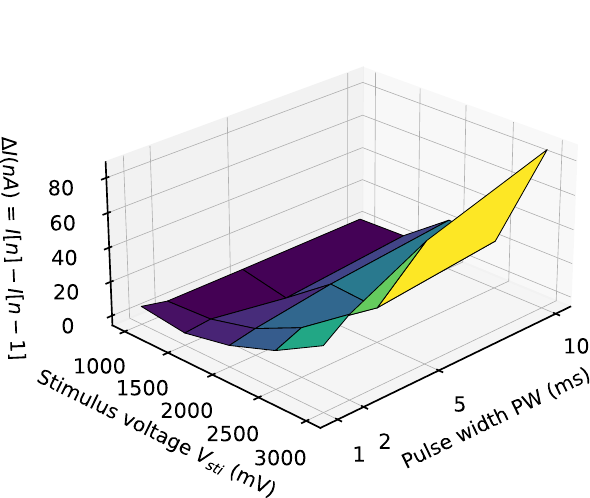}
% \caption{\textcolor{red}{Measured current change ($\Delta I$) surface of a volatile memristor.}}
\caption{Measured current-change ($\Delta I$) surface of the volatile HfTiO$_x$ memristor as a function of programming voltage and pulse width. The surface shows that larger stimulus amplitudes and longer pulses produce progressively higher $\Delta I$.}
\label{fig:gain surface}
\end{figure}

In our HfTiO$_x$ devices (detailed in Section \ref{sec:mem_fab}), we observe bi-exponential current decay on speech-relevant timescales (Fig.~\ref{fig:STP and RRAM}), consistent with the mechanisms and ranges reported above, enabling a return of the ADM threshold $\Delta$ without additional active control energy after a brief programming pulse.
Physiological recordings and standard auditory-nerve models show that spike-rate adaptation requires at least two exponential components: a rapid component of a few milliseconds and a slower component in the tens of milliseconds\cite{smith1985origins}. Adopting a bi-exponential decay for the spike threshold therefore mirrors this more realistic auditory-periphery behaviour, capturing both the sharp onset drop and the slower recovery in firing rate.

The gain for a delta modulation system is typically defined as the reciprocal of the threshold ($\Delta$) at the encoder side\cite{gouveia2010asynchronous}. The intuition is that, for a given signal, a smaller $\Delta$ yields a greater response and a higher spike count.
Therefore, we can define the onset enhancement gain for a delta modulation system as the ratio of the post-onset threshold and the pre-onset threshold, which is proportional to the current ratio if the memristor current is designed to control the threshold as in this paper.
In other words, it is defined in terms of the device current after a stimulus ($I[n]$) and before the stimulus ($I[n-1]$), as expressed in Eq.\ref{eq:adm gain}.

\begin{align}
    \label{eq:adm gain}
    \rho_{ADM} &= \frac{I[n]}{I[n-1]} =\frac{I[n-1] + \Delta I}{I[n-1]} \notag \\
               &=  1 + \frac{\Delta I}{I[n-1]}
\end{align}

\begin{figure}[h]
\centering
\includegraphics[width=0.45\textwidth]{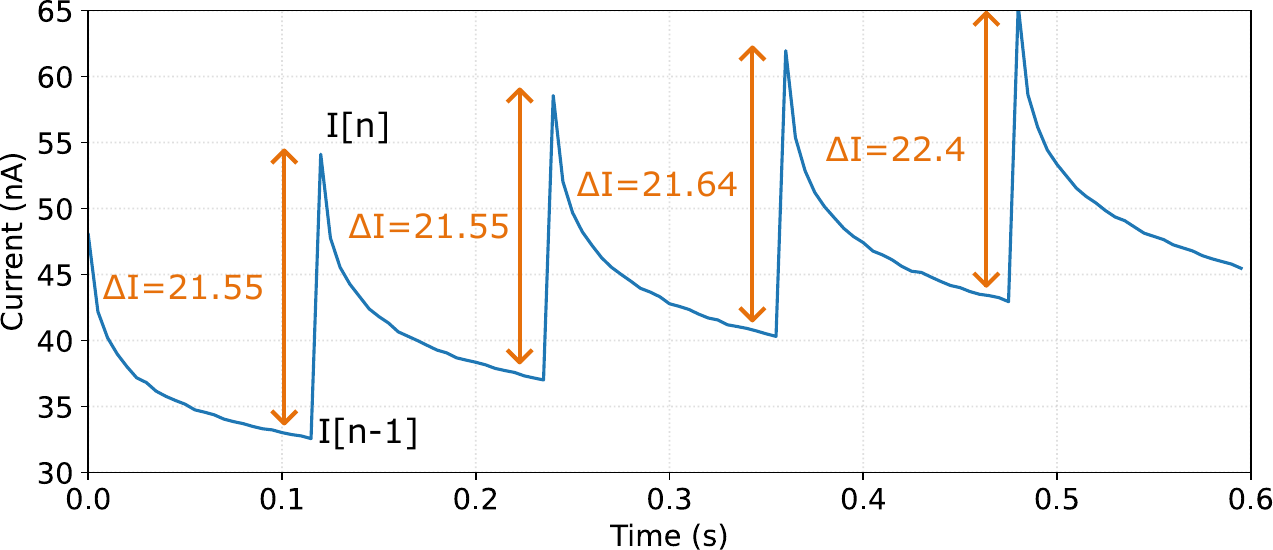}
% \caption{\textcolor{red}{Current change ($\Delta I$) for 5 consecutive stimuli (2.5V 5ms).}}
\caption{Current change ($\Delta I$) for five consecutive 2.5~V, 5~ms stimuli applied to the same volatile memristor.}
\label{fig:current_5T}
\end{figure}

A current change ($\Delta I$) surface was derived from the measurement result, and it shows the relation between the current change of the device and the stimulus pattern (of a single voltage pulse). It's interesting to note that the current change depends primarily on the stimulus voltage and its pulse width (as shown in Fig.\ref{fig:current_5T}).
Although the gain expression is related to another parameter $I[n-1]$, apart from the stimulus voltage ($V_{sti}$) and pulse width (PW), $I[n-1]$ actually would iterate with each stimulus so it can usually be modulated to a custom level.
Therefore, variation in the initial device resistance typically does not pose a significant problem in this application.

It is worth noting that the small variation of the current change surface will be acceptable as it can be compensated either by 1) tuning the stimulus parameters ($V_{sti}$ and PW) or 2) tuning the gain of the preceding stage as there will typically be gain stages within the conditioning circuits in auditory front-ends. Also, future volatile devices with a much steeper current-change surface (switchable with lower voltage and shorter pulse width) would further benefit this application.

Beyond the qualitative match in dynamics, volatile memristors also offer tangible implementation benefits over CMOS-only realisations of stimulus-dependent decay. 
For example, Alea \emph{et al.} \cite{alea2024fingertip} implement millisecond-scale decay using an on-chip capacitor and explicitly note that a significant portion of the taxel footprint is dictated by this large capacitor. 
From the layout, the capacitor is estimated to occupy about \(75 \times 75~\mu\text{m}^2\) per channel, whereas the volatile HfTiO\(_x\) devices used in our prototype have an active area of \(20 \times 20~\mu\text{m}^2\). 
Replacing a large on-chip capacitor with a volatile memristive device therefore has the potential to reduce the area of the adaptive element by more than an order of magnitude (92.9\%) and to ease scaling to dense multi-channel arrays. 
Moreover, accessing the decay using material physics through a small read bias leads to nanoampere-level current consumption, suggesting a path to low static power for the adaptation mechanism, even though a full power comparison with CMOS-only implementations is beyond the scope of this work.

Considering the similarity between the memristor's current decay and short-term adaptation, it is a natural fit to implement a short-term adaptation-like control mechanism
if the current of the device can control the response strength of a spike encoder.

%% file: contents/system_architecture.tex
\section{System Architecture}

\begin{figure*}[!ht]
\centering
\includegraphics[width=0.9\textwidth]{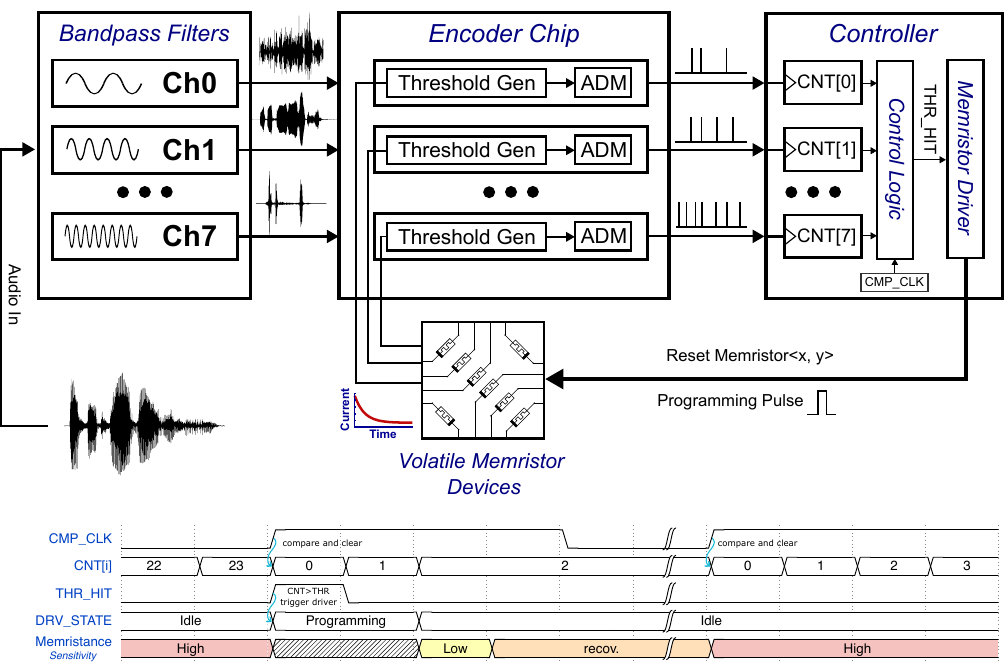}
\caption{(top) System architecture of the proposed system; (bottom) timing diagram of a single channel in the system.}
\label{fig:system arch}
\end{figure*}

The proposed memristor-assisted system is shown in Fig.~\ref{fig:system arch} (top). The system consists of
a bandpass filter bank, a custom encoder chip (implemented in standard 130 nm technology and reported in \cite{guo2025multi}), a logic controller, and a volatile memristor array.

The raw input sound first passes through the bandpass filters, where components of different centre frequencies
are decomposed in a manner inspired by the cochlea. The bandpassed signals are then fed to the encoder chip, which consists of multiple channels of
asynchronous delta modulators (ADMs). Each ADM's threshold ($\Delta$) is controlled by the resistance of a volatile memristor (memristance) within
the array.

The logic controller monitors the real-time spike output from all channels and programs the corresponding device when a specific channel is too active,
to modulate its activity and emulate the adaptive response found in short-term adaptation.

The bottom panel of Fig.~\ref{fig:system arch} shows the timing diagram illustrating how the system adjusts its threshold for a single channel. 
An asynchronous counter (CNT[i]) counts the spikes generated by the ADM and is periodically evaluated and reset by a clock signal CMP\_CLK. On the
first rising edge of CMP\_CLK, CNT[i] (the value of the ith counter) is read. When the spike count exceeds the threshold for \textit{high activity} (THR, set to 20 in this example), the THR\_HIT signal is asserted.

When THR\_HIT is asserted, the memristor driver activates to send a programming pulse to the corresponding device. Before programming, the device is disconnected from the encoder chip, and
the last available threshold is sampled and held to allow the ADM to operate without disruption (more details in Section \ref{sec:RRAM Interface}). The programming pulse lowers the device's memristance, which is translated into a higher threshold (\textit{lower sensitivity}) for spike conversion
by the on-chip threshold generator (detailed in Section \ref{sec:circuit}). As a result, spikes are generated less frequently and the spike rate drops.

The inherent volatility of the memristor causes its memristance to automatically return to a high-resistance state without any external stimulus.
After a period of tens of milliseconds, the threshold gradually decreases to its initial value, restoring the system's sensitivity for detecting new input sounds.
This dynamic modulation of spike activity mimics the short-term adaptation observed in biological systems, effectively enhancing the encoding of temporal features in audio signals while reducing the overall spike count.

%% file: contents/implementation.tex
\section{System Design and Implementation}

\subsection{Spike Encoder Chip}
\label{sec:circuit}
\begin{figure}[h]
\centering
\includegraphics[width=0.45\textwidth]{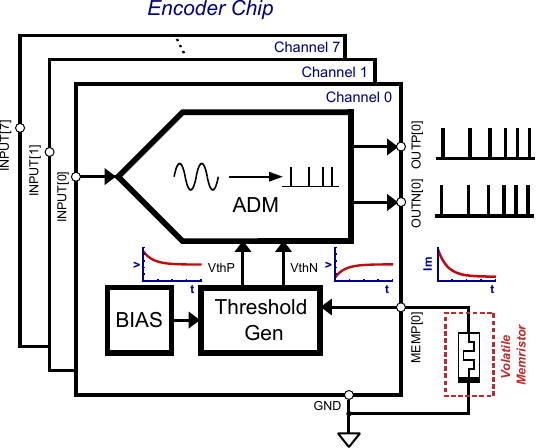}
\caption{Architecture of the Spike Encoder Chip}
\label{fig:circuit arch}
\end{figure}

There are eight channels in the encoder chip; each channel consists of a threshold generator, an asynchronous
delta modulator (ADM), and a local bias module as shown in Fig.\ref{fig:circuit arch}. The threshold generator is connected to an off-chip memristor device, whose current is translated into a threshold voltage ($\Delta$) for the ADM. 
The ADM converts the analogue input signal to digital spikes based on the principle of delta modulation, with the dynamic threshold controlled by the memristor.

\subsubsection{Threshold Generator}

\begin{figure}[h]
\centering
\includegraphics[width=0.4\textwidth]{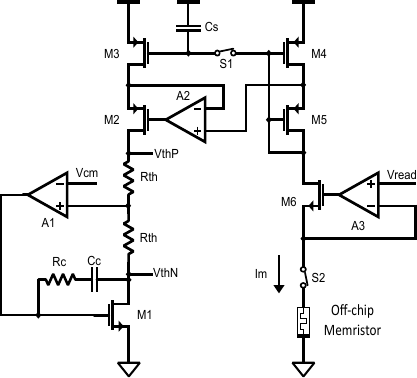}
\caption{Schematic of the Threshold Generator\cite{guo2025multi}}
\label{fig:threshold generator}
\end{figure}

The threshold generator converts the memristor current into a pair of symmetric comparator thresholds (VthP, VthN) around Vcm, which define the ADM step size $\Delta$ (Fig.~\ref{fig:threshold generator}).

To generate symmetrical threshold voltages utilising the dynamic current (Im) of the memristor, 
a current-mirror-based transimpedance amplifier is designed as shown in Fig.\ref{fig:threshold generator}.
The circuit receives the current from the memristor (Im) in the right branch of the current mirror and then copies and amplifies it to the left branch; the amplified memristor current then passes through two resistors (Rth) and is converted into threshold voltages (VthP and VthN).

A regulated read voltage (Vread) is applied to one end of the device (MEMP[i]) through a current follower (consisting of A3 and M6), to keep the voltage drop constant during normal operation, as the memristor behaves like a diode under DC conditions.

The resulting current (Im=Vread/Rmem) is then copied and amplified by a gain-boosted cascode current mirror ($\beta$ = M3:M4=10:1), as the memristor's current is very small. The amplified current then
flows through two matched resistors, generating symmetrical voltages centred around the common-mode voltage (Vcm). The central voltage is regulated by A1 and M1.

The gain of this circuit is defined by $\beta \times \mathrm{R_{th}}$, where $\mathrm{R_{th}}$ is the resistance of the two resistors. $\mathrm{R_{th}}$ is designed to be 30 k$\Omega$, as a trade-off between area and circuit stability.  
The overall transimpedance gain ($\mathrm{G_{TIA}}$) is around 300 k$\Omega$, which is sufficient to convert the nanoamp-range current of the memristor to a millivolt-range voltage.

A sample and hold (S\&H) circuit consisting of S1 and Cs is added between the two branches to hold the bias voltage for M3 when the memristor is disconnected for programming. 
One end of the hold capacitor is connected to VDD for better Power Supply Rejection Ratio (PSRR).

The OTA used for regulating the mid-point voltage (A1) and the transistor used as a variable resistor (M1) form a
closed-loop two-stage amplifier; therefore, a feedforward resistor (Rc) and a capacitor (Cc) are added
to compensate for loop stability.

\subsubsection{Asynchronous Delta Modulator}

\begin{figure}[h]
\centering
\includegraphics[width=0.47\textwidth]{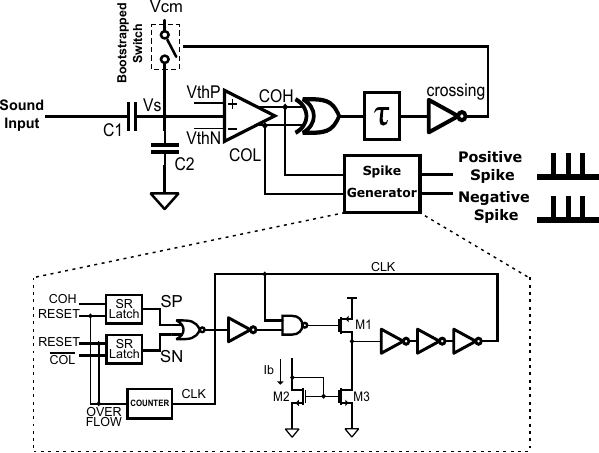}
\caption{Schematic of the Asynchronous Delta Modulator\cite{guo2025multi}}
\label{fig:ADM}
\end{figure}

The ADM converts the analogue input into an asynchronous stream of spikes such that each spike corresponds to a fixed voltage increment $\Delta$ at the input node (Fig.~\ref{fig:ADM}).

A feedback-and-reset loop\cite{yang2014comparison} is designed to implement the asynchronous delta modulation, as shown in Fig.\ref{fig:ADM}, 
since only the instantaneous spike needs to be generated, not the digital representation of the signal.

The analogue input is fed to a capacitive divider consisting of C1 and C2 (C1:C2=3:1).
Then a three-level comparator\cite{hou20181} compares the node voltage Vs with two threshold voltages (VthP and VthN) provided by
the threshold generator. Whenever Vs exceeds the boundary defined as [VthN, VthP], the succeeding logic gates
will trigger the bootstrapped switch to bring the node voltage back to Vcm, and a short pulse at the comparator output (COH or COL) is generated, depending on which boundary is crossed. Then the comparator continues to monitor the node voltage until it crosses the threshold again;
hence, the delta modulation is performed asynchronously.

Considering the timing of this asynchronous process, a delay cell ($\tau$)
is added in the path to ensure the reset signal (\textit{crossing}) holds long enough for Vs to settle when Vs exceeds the boundary. 
To trade off the modulation timing error and the bootstrapped switch's speed, 
the delay cell is designed to have a delay of approximately 10 ns, which is sufficient for the node voltage to settle and will only cause a maximum timing error (jitter) of 10 ns in the spike output.

Despite keeping the timing error small, such a short pulse width of the spikes imposes strict constraints on the speed of the
following back-end system. Therefore, a spike generator based on \cite{timmermans20241} that generates spikes
with adjustable pulse width is added in parallel with the main loop. 

The short pulses COH and COL generated by the main loop set a pair of SR latches and enable a ring-oscillator-based clock. A counter then counts clock cycles until overflow, at which point the SR latches are reset. The extended spike outputs SP and SN are taken from the latch outputs, so each short pulse is stretched to a programmable duration. The pulse width is set by the current through M2 (which determines the frequency of the oscillator), allowing adjustment from 10~ns to 500~ns using an off-chip 5~M$\Omega$ trimmer.

\subsection{Memristor Interface}
\label{sec:RRAM Interface}
The memristor interface allows the controller to temporarily disconnect each device from the encoder chip and connect it to an external pulse generator for programming, while preserving the on-chip bias conditions used by the threshold generator.

For rapid development and a quick proof of concept, we used a commercially available electronic-device testing platform (ArC TWO~\cite{foster2022fpga}) as the off-chip pulse driver to program the memristor device. ArC TWO provides 64 independent source–measure units (SMUs) that can be configured for read or write operations. For write operations, the platform supports programmable pulse amplitudes up to \(\pm 13.5~\mathrm{V}\) with pulse widths down to tens of nanoseconds. In our system, the controller issues triggers to ArC TWO to deliver brief pulses that transiently increase device conductance; the subsequent recovery is provided by the intrinsic volatility of the devices, so no explicit RESET pulses are required.
\begin{figure}[h]
\centering
\includegraphics[width=0.45\textwidth]{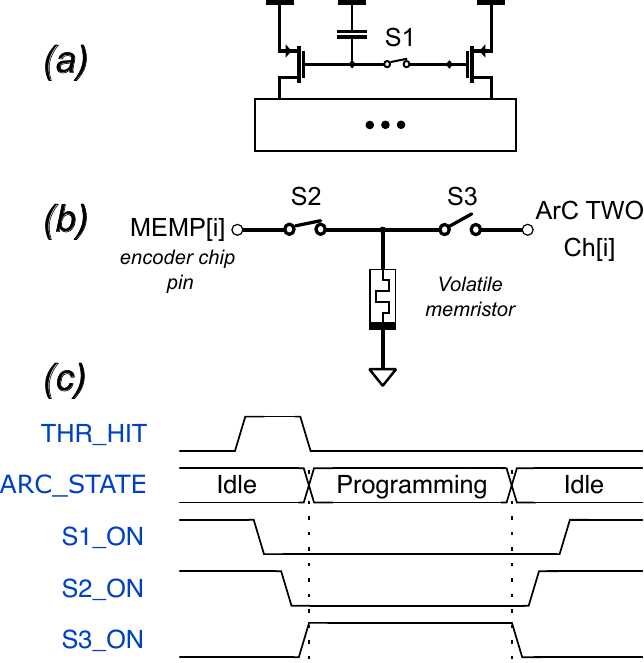}
\caption{Memristor interface switching; (a) switch on the encoder chip; (b) switches between encoder chip and ArC TWO; (c) timing diagram of the switching.}
\label{fig:mem_interface}
\end{figure}

After being triggered, the sample-and-hold (S\&H) switch on the encoder chip is first opened, as shown in Fig.\ref{fig:mem_interface}.
Then switch S2 is opened to disconnect the device from the chip. Finally, S3 is closed to connect the device to ArC TWO channels for
the programming. After programming, the switches are manipulated in reverse order - open S3, close S2, then S1, to ensure that the on-chip threshold generator is not affected during this procedure. All these switching controls are performed by the controller.

The programming applies a 5 ms, 2.5 V pulse to the device through a DAC followed by driver circuits on ArC TWO. The pulse width and voltage are programmable and represent a trade-off between speed and power consumption.

\subsection{Memristor Fabrication}
\label{sec:mem_fab}
Thin-film memristive devices with an active area of $20 \times 20~\mu\text{m}^2$ were fabricated at room temperature. The top and bottom TiN electrodes were deposited by reactive sputtering using an Angstrom system in a nitrogen ambient at a working pressure of 2~mTorr. An 8-nm HfTiO$_x$ switching layer was grown using atomic layer deposition (ALD) at $250^{\circ}$C. Tetrakis (dimethylamido) hafnium (IV) (TDMAHf) and tetrakis (dimethylamido) titanium (TDMATi) were used as the Hf and Ti precursors, with water serving as the oxidant. Electrical characterisation was carried out using a Keithley 4200 Semiconductor Characterization System (SCS) to study the volatile behaviour. A die containing 32 devices was subsequently packaged in a PLCC68 package.

\subsection{Controller}
An Arm Cortex-M7 processor is used to implement the controller for fast development. The on-chip timers are configured in external clock-source mode to count the spikes from the encoder chip with minimal delay. 
Another timer is set to generate a 100 Hz comparison clock (CMP\_CLK) to regularly check and reset the counter values in an interrupt routine. When the counter value is higher than a pre-defined value (THR),
the spike activity is deemed too high, and the switching followed by the programming procedure of the corresponding device is triggered.

\subsection{Parameter Selection}

The key parameters governing the adaptive-thresholding behaviour are the comparison clock frequency (CMP\_CLK), the spike-count threshold (THR\_HIT), and the programming pulse amplitude (V\textsubscript{sti}) and width (pw).

CMP\_CLK was set to 100~Hz, corresponding to a 10~ms update period. This defines the temporal resolution of the adaptation response: it must be sufficiently fast compared to the audio dynamics (i.e., shorter than a typical speech syllable) to ensure a timely response, while a much higher rate would increase computational load and energy consumption. A value of 100~Hz provides a practical compromise between responsiveness and overhead.

The spike-count threshold THR\_HIT determines when a channel is considered to be overactive and thus requires reset. Its choice reflects a trade-off between robustness to noise (avoiding false triggers due to spurious spikes) and sensitivity to genuine increases in spike rate. In the prototype, THR\_HIT was set to 50, which we found to strike a reasonable balance between these two requirements.

V\textsubscript{sti} and pw define the stimulus applied to the memristor during programming. Both must be large enough to induce a noticeable increase in device conductance (more details in Fig.\ref{fig:gain surface}) and thereby produce a meaningful onset gain enhancement, but larger values also increase programming energy. In addition, pw directly determines how long the device is disconnected from the encoder chip; this interval should be kept as short as possible to minimise disruption to normal operation. We therefore chose a pulse width of 5~ms as a compromise between programming efficacy and interruption time. The parameter values used here are not necessarily optimal, and the system offers a wider parameter space that can be exploited to optimise performance for different noise conditions and application scenarios.

%% file: contents/measurement_results.tex
\section{Hardware Measurement Results}

A prototype chip consisting of 8 channels of the encoding circuit was fabricated in a standard 130~nm CMOS process. 
The functionality of each sub-circuit was first verified with a custom testboard, which was designed to
provide the necessary biasing and reference voltages. The complete system, as illustrated in Fig.\ref{fig:system arch} (with the gammatone filter bank implemented using off-line software filters) was then tested with all the modules connected together (setup shown in Fig.\ref{fig:system setup}).

\subsection{CMOS Chip Measurement}
\label{subsec:CMOS_result}
\subsubsection{Threshold Generator}

\begin{figure}[h]
\centering
\includegraphics[width=0.48\textwidth]{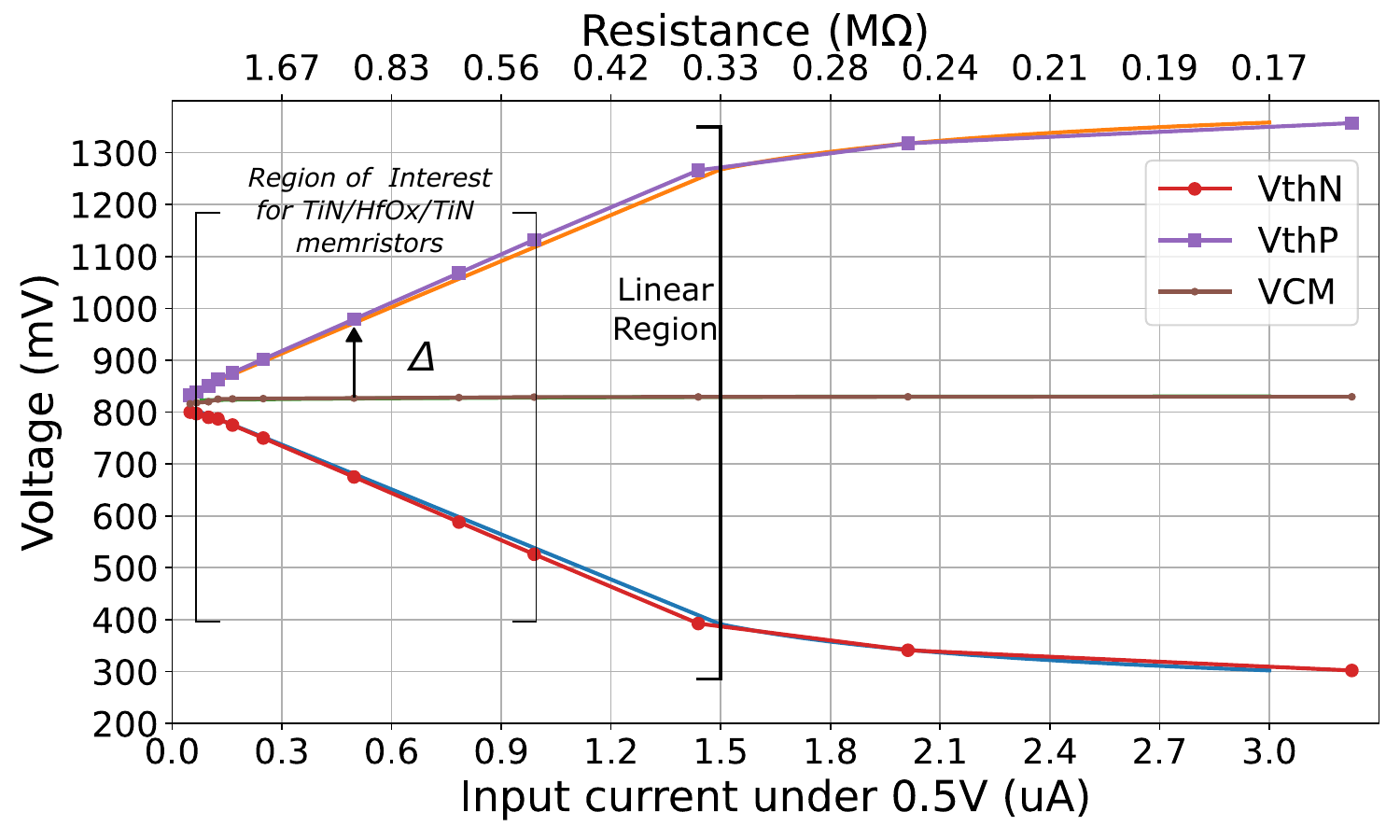}
\caption{Measured transfer function of the threshold generator}
\label{fig:threshold generator measurement}
\end{figure}

The transfer function of the threshold generator (as shown in Fig.\ref{fig:threshold generator measurement}) was characterised using a trimmer resistor on the testboard that covers the resistance range of the volatile memristor eventually used in the system. By sweeping the resistance from 1.5 M$\Omega$ to 100 k$\Omega$, the resulting input current to the threshold generator varies from 0.066 $\mu$A to 0.99 $\mu$A (input current is derived from $\mathrm{V_{read}}/\mathrm{R_{trimmer}}$, where $\mathrm{V_{read}}$ is the regulated read voltage across the trimmer and is set to 0.5 V). Both the resistance of the trimmer and the output voltage are measured with a 6½-digit digital multimeter (Agilent 34410A).

The figure shows that the measurement results (dotted line) align well with the simulation results (solid line). 
The linear range of the sub-circuit is 0-1.5 $\mu$A, which is sufficient to accommodate the current range of different memristor devices. 
The symmetrical reference voltage is suitable for defining the delta modulation threshold $\Delta$ (VthP = VCM + $\Delta$, VthN = VCM - $\Delta$),
which could have a tuning range as large as 5 mV to 400 mV, subject to the input current range.

\subsubsection{Asynchronous Delta Modulator}
\begin{figure}[h]
\centering
\includegraphics[width=0.489\textwidth]{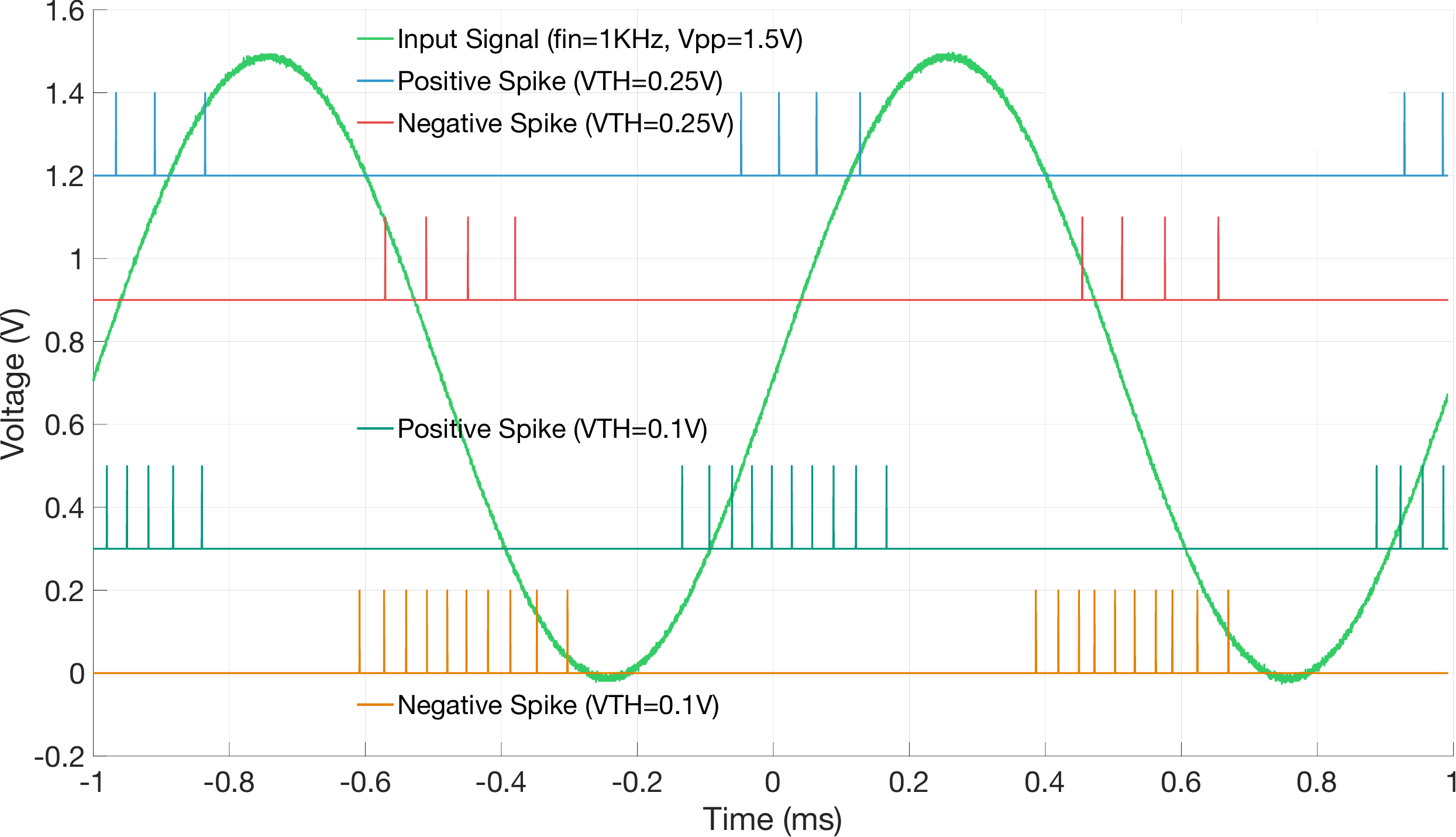}
% \caption{Spike output of the ADM, under different threshold voltages ($\Delta = 0.1/0.25V$)}
\caption{Spike output of the ADM under two threshold voltages, $\Delta = 0.1~\text{V}$ and $0.25~\text{V}$. The spike train is noticeably denser for the smaller threshold ($0.1~\text{V}$), illustrating the higher encoder response when $\Delta$ is reduced.}
\label{fig:ADM measurement}
\end{figure}

The Asynchronous Delta Modulator (ADM) was tested with a 1 kHz full-scale (0-1.5 V) sine wave, with a fixed
threshold voltage generated by on-board voltage dividers. The output was captured as raw waveforms with a digital oscilloscope (R\&S RTO2064).

\begin{figure}[h]
\centering
\includegraphics[width=0.475\textwidth]{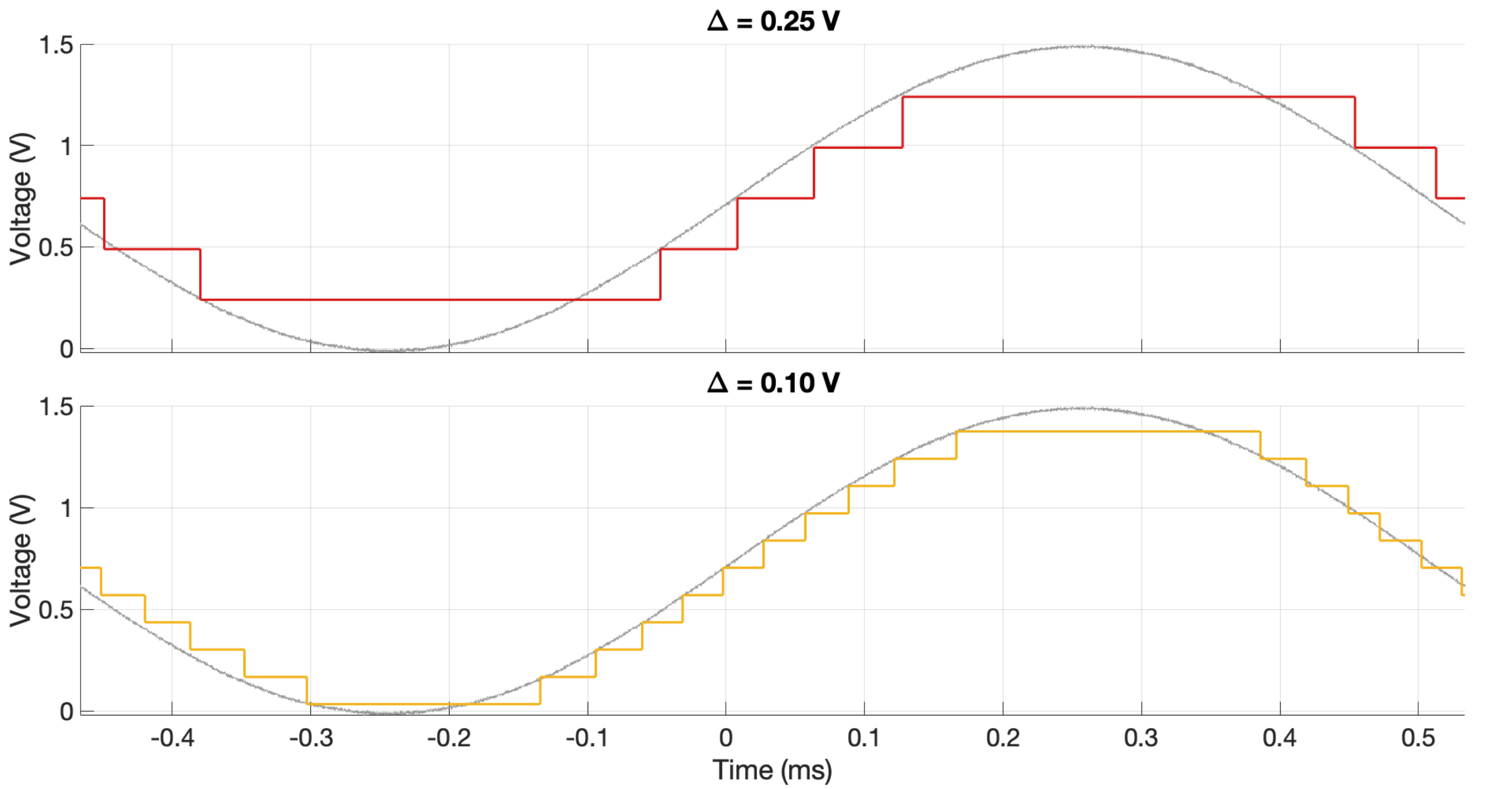}
% \caption{Reconstructed output of the ADM}
\caption{Reconstructed output of the ADM for the same 1~kHz full-scale (0--1.5~V) sine-wave input used in Fig.~\ref{fig:ADM measurement}. The reconstruction is obtained by integrating the recorded spikes and scaling by the threshold $\Delta$, producing a waveform that closely tracks the original amplitude and phase of the input.}
\label{fig:ADM reconstruction}
\end{figure}

As shown in Fig.\ref{fig:ADM measurement},
the ADM responds differently for different threshold voltages ($\Delta$); the spikes are denser when the input has a larger derivative, and also denser for smaller threshold voltages.
From the recorded spike output, a simple reconstruction can be performed to validate the correct timing of spike generation.
The spike output is integrated over one period of the input sine wave, and the output is reconstructed by summing the spikes and scaling them according to the threshold voltage ($\Delta$). 
The reconstructed output is shown in Fig.\ref{fig:ADM reconstruction}.

It can be seen that the reconstructed output is a good approximation of the input sine wave, and the ADM
works well with different threshold voltages ($\Delta$). 

\begin{figure}[h]
\centering
\includegraphics[width=0.475\textwidth]{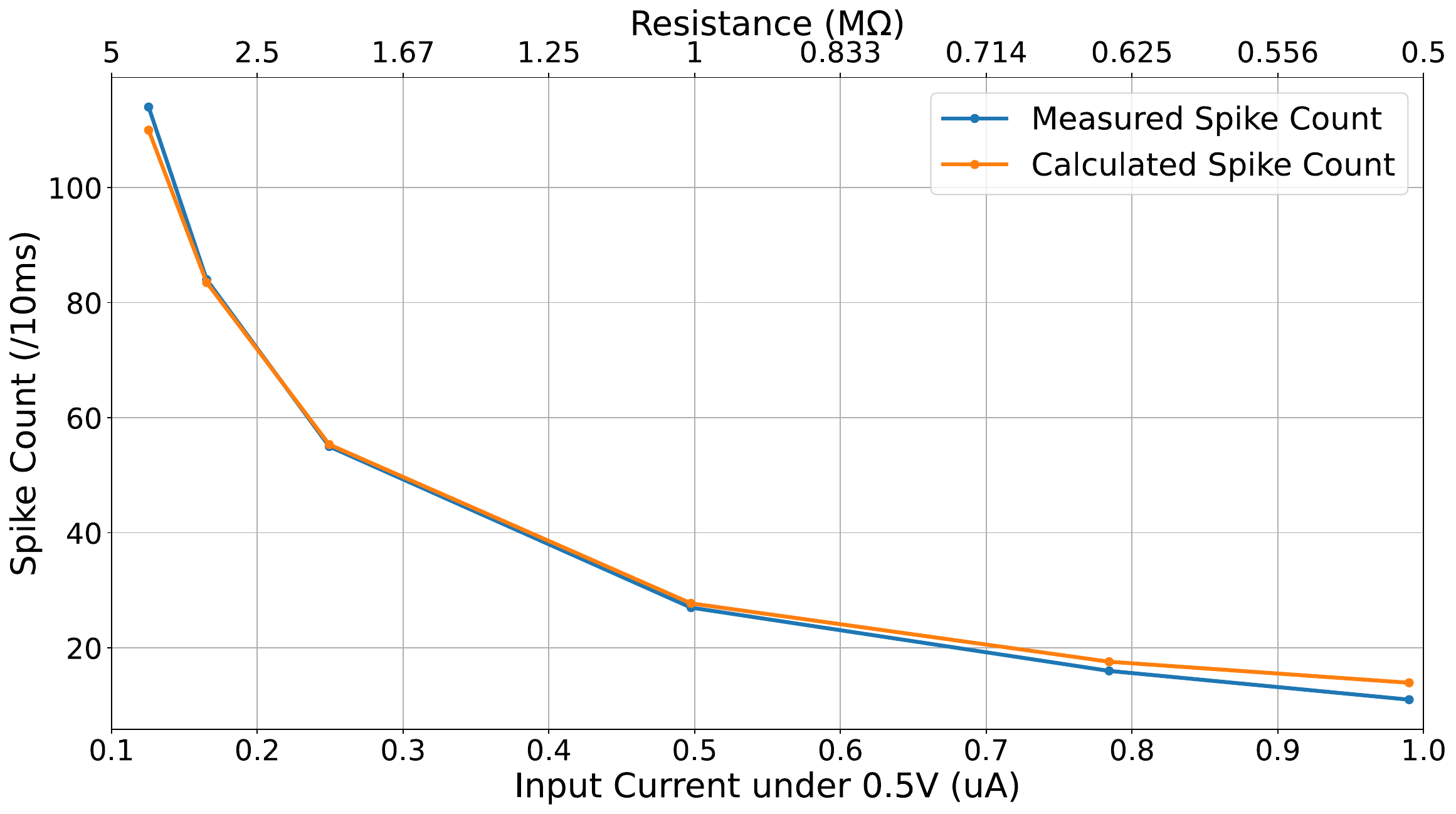}
\caption{Spike rate vs Input current, with a 200Hz full-scale sinusoidal signal}
\label{fig:Spike Rate}
\end{figure}

The spike-rate response of the ADM was also tested with a 200 Hz full-scale sine wave. The threshold voltage was provided by the Threshold Generator
and changed by tuning the off-chip resistance trimmer. 
When the trimmer was swept (hence the threshold $\Delta$), the spike count was measured and plotted in Fig.\ref{fig:Spike Rate}.
The spike count was obtained by counting the number of spikes (both SP and SN) in a 10 ms window. With a pure tone input $A\cdot \sin(2\pi f_{in} t)$, 
the spike rate of an ideal delta modulator can be expressed as:

\begin{equation}
\label{eq:spike_rate}
\begin{aligned}
f_{spike} &= 4 \cdot \frac{A \cdot f_{in}}{\Delta} \\
          &= 4 \cdot \frac{A \cdot f_{in}}{I_{in} \cdot G_{\mathrm{TIA}}}
\end{aligned}
\end{equation}

The spike rate is inversely proportional to the threshold voltage ($\Delta$) and directly proportional to the input amplitude ($A$) and frequency ($f_{in}$).
The measured spike count aligns well with the theoretical prediction (with Eq.\ref{eq:spike_rate}), as shown in Fig.\ref{fig:Spike Rate}, with minor discrepancies likely due to reset error when the threshold is too wide.
However, due to the use of high-resistance devices, the real resistance of the memristor is well controlled above 0.6 M$\Omega$, and this discrepancy will not significantly affect the overall performance.

\subsection{Memristor Characterisation}
\label{sec:mem_char}

\begin{figure}[h]
\centering
\includegraphics[width=0.485\textwidth]{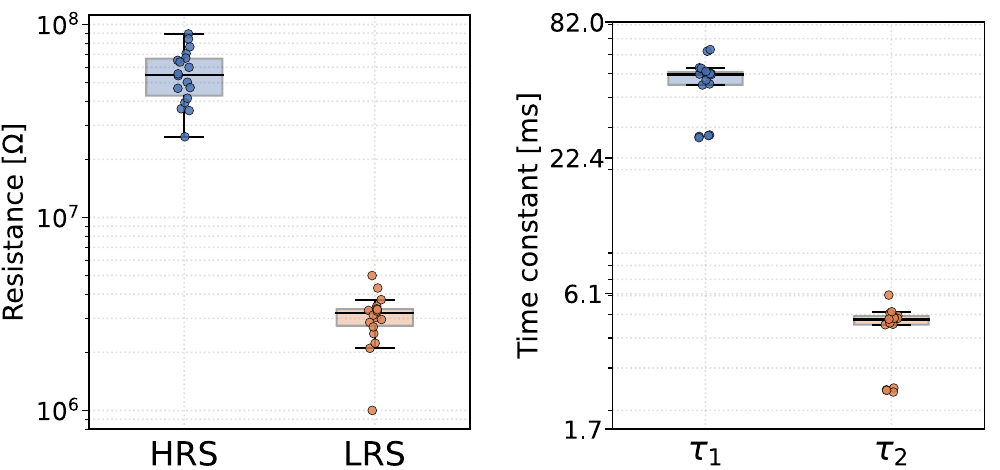}
%\caption{\textcolor{red}{Device variation of the 18 devices on the packaged die; (left) resistance distribution; (right) time constant distribution;}}
\caption{Device variation of the 18 devices on the packaged die; (left) resistance distribution (HRS: initial state; LRS: after three 3~V, 10~ms pulses); (right) distributions of the slow ($\tau_1$) and fast ($\tau_2$) decay time constants, extracted from current transients following a 2.5~V, 5~ms pulse, highlighting the large resistance window and tight spread in device dynamics across all devices.}
\label{fig:mem_dis}
\end{figure}

Figure~\ref{fig:mem_dis} summarises the electrical characterisation of the 18 memristive devices on the packaged die. The left panel shows the distribution of the high-resistance state (HRS) and low-resistance state (LRS) across all devices. The HRS is defined as the initial resistance measured before any external programming stimulus is applied, and therefore reflects the as-fabricated state of the devices. The LRS is defined after the application of three consecutive 3V 10ms programming pulses, providing a rough indication of the low resistance levels that are relevant for our intended operation. The resulting HRS–LRS resistance range is relatively large, which is advantageous because it allows us to reliably tune the device conductance during the initialisation phase so that each memristor can be matched to the requirements of the surrounding analogue circuitry.

The right panel of Fig.~\ref{fig:mem_dis} shows the distributions of the two characteristic time constants, ($\tau_{1}$) and ($\tau_{2}$), extracted from the transient response when a 2.5V 5ms pulse is applied. Both ($\tau_{1}$) and ($\tau_{2}$) exhibit a comparatively narrow spread across the 18 devices, and their absolute values fall within a range that is well suited for the temporal dynamics required in audio-processing tasks\cite{delgutte1980representation}. Consequently, device-to-device variability in the dynamical properties is small enough that no additional compensation or elaborate calibration is required at the system level, and memristor variation does not pose a limitation for the intended application.

% \begin{figure}[h]
% \centering
% \includegraphics[width=0.475\textwidth]{figs/mem_char_protocal.pdf}
% \caption{}
% \label{fig:mem_protocal}
% \end{figure}

\subsection{Encoder System Measurement}
\subsubsection{Setup}
\begin{figure}[h]
\centering
\includegraphics[width=0.48\textwidth]{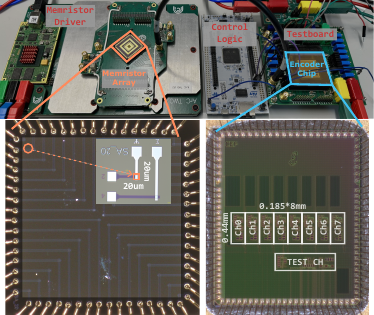}
\caption{Setup of the encoder system; (bottom left) Micrograph of the memristor devices; (bottom right) Micrograph of the encoder chip;}
\label{fig:system setup}
\end{figure}

The whole encoder system (as illustrated in Fig.\ref{fig:system arch}) was set up as shown in Fig.\ref{fig:system setup}.

The system comprises three main components: the Memristor Driver, implemented using an ArC TWO board; the Controller, based on an STM32H753ZI microcontroller; and the testboard, which houses the Encoder Chip.

HfTiO$_\text{x}$ memristors were fabricated in-house, each $20 \mu$m $\times$ $20 \mu$m. A die consisting of 32 devices was then packaged into PLCC68 and plugged
into the daughterboard of ArC TWO. 

The encoder chip was fabricated using standard 130 nm technology. It consists of 8 general-purpose channels, and each channel occupies an area of
0.44 mm $\times$ 0.185 mm. A dedicated test channel is included to enable comprehensive characterization of the results, as detailed in Section \ref{subsec:CMOS_result}.

The connection from the memristor device to the encoder chip is made with an SMA cable between the daughterboard and the testboard, with analogue switches in between to disconnect the device from the Encoder Chip when it needs programming (as described in Section \ref{sec:RRAM Interface}).

The output of the encoder chip is connected to the Controller to monitor the spike rate in real time.
The timer of the microcontroller is configured in external clock mode, which counts the number of spikes in a pre-defined time window.
When the real-time spike count is above a certain threshold, the microcontroller will trigger the ArC TWO board
to apply a short pulse (with custom width and amplitude) to the memristor device, which is connected to the input of the Threshold Generator on the encoder chip.

\subsubsection{Speech Dataset Preparation}
EmoV-DB\cite{adigwe2018emotional} was used as the dataset for the encoder system measurement. Off-line band-pass filtering using a gammatone filter\cite{slaney1998auditory} was applied to the audio files
to mimic cochlear filtering in Matlab.
The filtered audio signals were then amplified (to 400 mVpp) and applied to the encoder chip as input with an arbitrary waveform generator (AWG) on the oscilloscope (R\&S RTO2064).

\begin{figure}[h]
\centering
\includegraphics[width=0.49\textwidth]{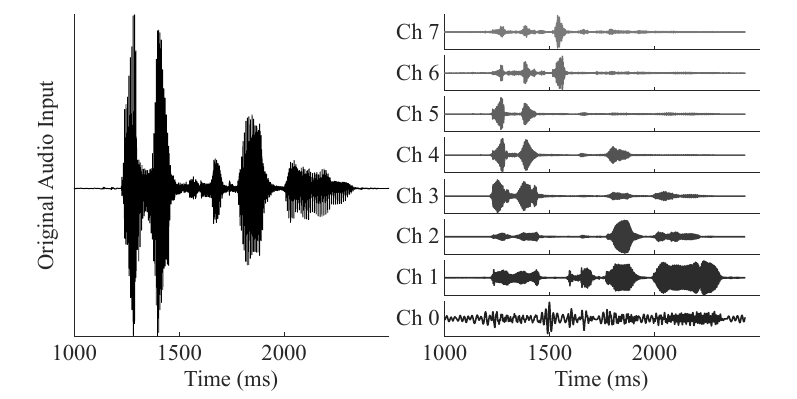}
\caption{Gammatone Filtering of the Audio Signal; (left) Original audio signal; (right) Filtered audio signal (all normalised to 400mVpp)}
\label{fig:gamma filtering}
\end{figure}

The test audio speech signal is shown in Fig.\ref{fig:gamma filtering}, which is a 1.5-second-long speech signal sampled at 16 kHz, with a man speaking \textit{'We have to be careful with them'} in a neutral tone.
The gammatone filter has a frequency range of 50 Hz (Ch0) to 8 kHz (Ch7), which is suitable for human speech signals.

\subsection{Single Channel Results}

\begin{figure}[h]
\centering
\includegraphics[width=0.49\textwidth]{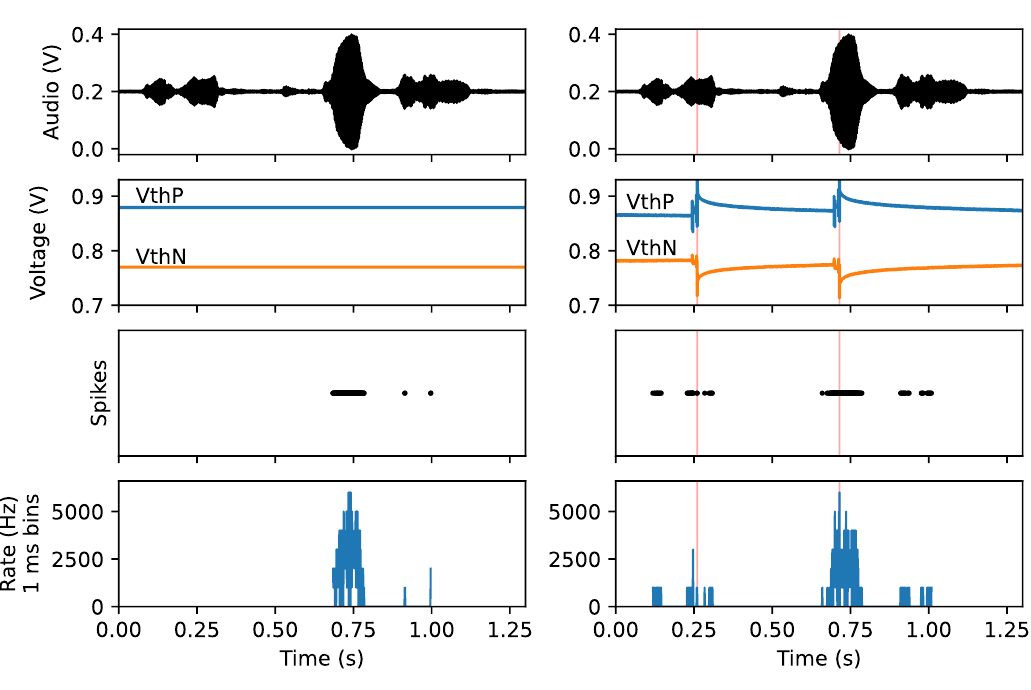}
%\caption{Single Channel Results; (left) Baseline Results; (right) Results with Adaptive Thresholding}
\caption{Single-channel encoding results for a 504.6~Hz band-pass filtered speech signal: the adaptive-threshold encoder (right) produces strong onset responses while suppressing steady-state spikes compared to the fixed-threshold baseline (left), improving representation of speech onsets under the same spike-count budget.}
\label{fig:Single Channel Results}
\end{figure}

The single-channel results were obtained by applying the filtered audio signals to a single channel of the encoder chip, and the spike output (only the positive channel) and the threshold voltage (VthP and VthN) were recorded with an oscilloscope.
The results are shown in Fig.\ref{fig:Single Channel Results}, where the left figure shows the baseline results with fixed threshold voltage, and the right figure shows the results with adaptive thresholding.
The baseline results were obtained through simulation, while the adaptive ones were from measurement. 
In the baseline case, an ideal delta modulator was used to generate the spikes, and the threshold voltage ($\Delta$) was set to 57 mV to make the total spike count the same as the adaptive case (248 spikes during the measured time).

An ideal delta modulator in simulation was used as the baseline, to make it straightforward to adjust the threshold $\Delta$. This ideal modulator is free of circuit non-idealities such as electronic noise and offsets, and therefore represents an upper bound on performance; using it as a reference is thus a fair and conservative comparison that does not overstate the benefits of the implemented hardware.

The first row of the figures shows the band-pass filtered audio signal (corresponding to Ch2 in Fig.\ref{fig:gamma filtering}, centre frequency = 504.6 Hz). The second row shows the threshold voltage—where in the baseline case it is fixed to 820 ± 57 mV (to make the total spike counts the same in both cases),
and in the adaptive case it is dynamically adjusted according to the spike activity. The red line in the right figure indicates that the memristor is being reset, which is triggered by the microcontroller when the spike count exceeds a certain threshold.
The third row shows the spike raster plot, where each dot represents a spike event. The last row shows the real-time spike rate, which is the number of spikes counted in a 1 ms window.

In the adaptive case, the feedback resets the memristor device after the spike burst caused by the speech onset, which means a larger threshold voltage is applied for the signal after the onset, and the spike rate is reduced to a lower level (as shown in the spike rate drop in the bottom right figure).
Due to the bi-exponential decay of the memristor device, the threshold voltage gradually decreases (0.25-0.75 s, between the two red vertical lines), preparing for the next speech onset with a more sensitive threshold.

We quantify encoding fidelity by computing the Pearson correlation coefficient between the instantaneous spike rate and the root-mean-square (RMS) envelope of the input signal. The Pearson coefficient is defined in Eq.~\ref{eq:p_co}, where 
$s_t$ is the spike rate at time $t$, $x_t$ is the RMS value of the input at time $t$, and overbars denote time averages. Pearson's $r$ measures the strength of the linear association between the two signals; here it compares the spike-rate envelope with the stimulus energy. 
Over the full recording, the baseline encoder yields $r=0.922$ (92.2\%), while the adaptive encoder yields $r=0.943$ (94.3\%), indicating a slightly stronger tracking of the input envelope in the adaptive case.
When the analysis is restricted to the first 0.5\,s (onset period), the adaptive encoder achieves $r=0.677$ (67.7\%); the baseline produces no spikes in this interval, and the important onset information is completely lost.

\begin{equation}
    \label{eq:p_co}
    r \,=\, \frac{\sum_t\big(s_t-\overline{s}\big)\big(x_t-\overline{x}\big)}{\sqrt{\sum_t\big(s_t-\overline{s}\big)^2\;\sum_t\big(x_t-\overline{x}\big)^2}}
\end{equation}

This dynamic thresholding mechanism allows the most sensitive response to speech onsets while keeping the spike rate at a lower level during the steady state of the speech signal. When operating under a spike-count budget, this mechanism can help reduce the total spike count while enhancing the response to speech onsets.
As a comparison, in the baseline case under the same spike count budget, the threshold voltage is fixed to a relatively large value (less sensitive), which results in the finer details of the speech signal being lost, as shown in the left figure of Fig.\ref{fig:Single Channel Results} from 0-0.25 s.

\subsection{Multi-Channel Results}

\begin{figure}
    \centering
    \includegraphics[width=0.49\textwidth]{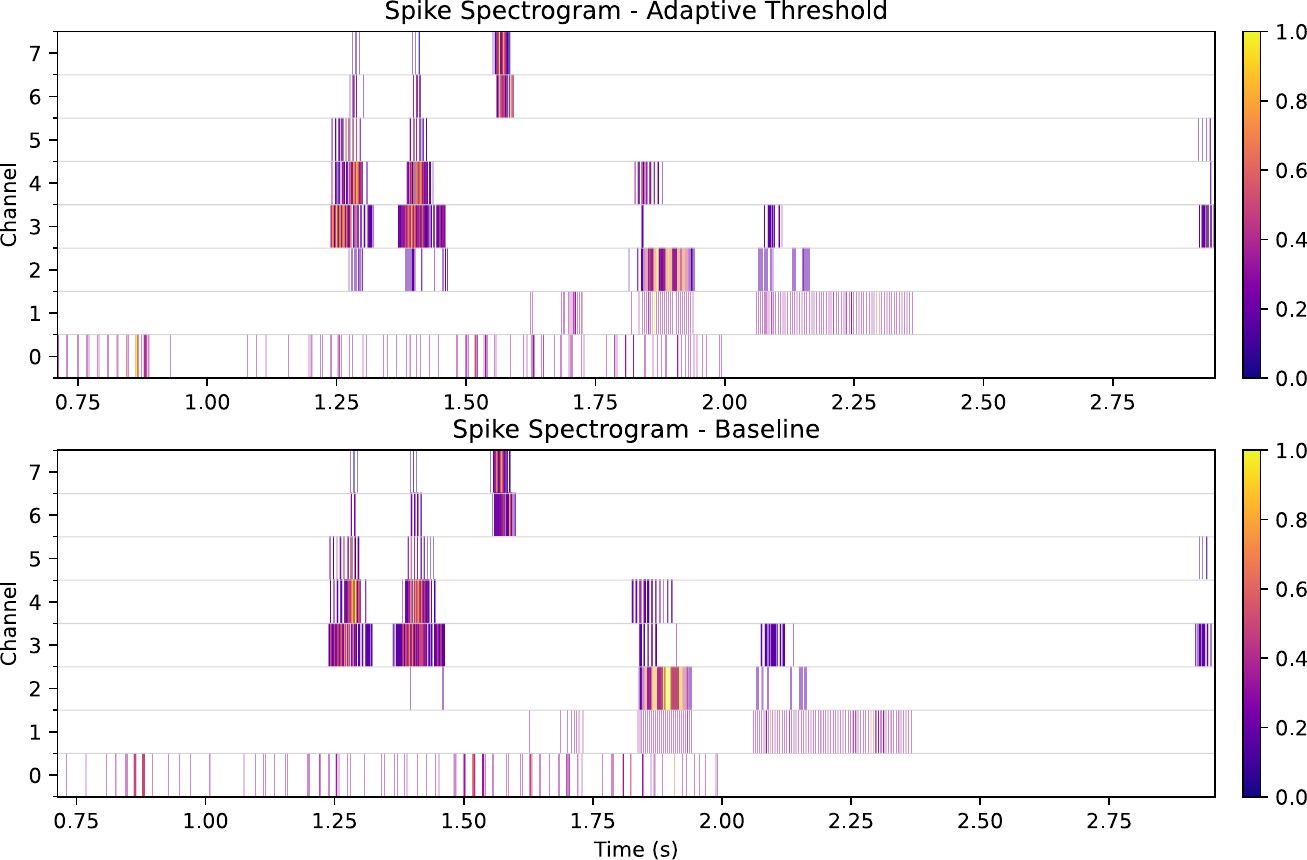}
    % \caption{\textcolor{red}{Multi-Channel Results}}
    \caption{Multi-channel cochleagram of the encoder output for an eight-channel gammatone-filtered speech signal spanning 50~Hz to 8~kHz; (top) the adaptive-threshold case, where spikes are concentrated at speech onsets while activity during steady segments is suppressed; (bottom) the fixed-threshold baseline tuned to yield the same total spike count as the adaptive case. Spike rates are normalised to 0--1 on a per-channel basis, highlighting the stronger onset contrast achieved with adaptive thresholding.}
    \label{fig:Multi Channel Results}
\end{figure}

With all eight channels of bandpass-filtered audio signals being fed into the encoder chip, 
multi-channel results can be obtained. The same measurement setup was used, and the spike outputs from all channels were recorded with the oscilloscope and then analysed. 
The results are shown in Fig.\ref{fig:Multi Channel Results}, where the top figure shows the cochleagram with the adaptive thresholding mechanism, and the bottom figure shows the cochleagram with fixed threshold voltage.
The fixed threshold was again adjusted to make the total spike count the same as the adaptive case.
The spike rate result is normalised to 0-1 channel by channel, and the colour bar indicates the intensity of the spikes.

From the results, it can be seen that the adaptive thresholding mechanism allows the encoder to respond effectively to speech onsets, exhibiting stronger spike activity at onsets compared to the steady state.
Also, when under the same spike count budget, the fixed threshold voltage results in poor representation of the speech signal, as the finer details are lost. In contrast, the adaptive case maintains these details by dynamically adjusting the threshold voltage and
represents the information content more effectively.

%% file: contents/conclusion.tex
\section{Conclusion}
This work has demonstrated and \emph{experimentally validated} an end-to-end \emph{hybrid CMOS--memristor} system 
for adaptive-threshold ADM-based spike encoding that emulates short-term auditory adaptation using the inherent volatility of HfTiO$_x$ devices. 
A brief program pulse increases \(\Delta\) after excessive spike activity, and the device’s bi-exponential relaxation 
then restores sensitivity autonomously—yielding onset-salient encoding without inflating the total spike budget. 
The system integrates an 8-channel 130\,nm encoder IC, off-chip memristor devices and a microcontroller 
that counts spikes and issues programming events through a switch interface. The on-chip threshold generator 
supports a wide \(\Delta\) range, enabling both sensitive and conservative regimes. 
Measured ADM spike-rate results follow the expected \(4A f_{\text{in}}/\Delta\) law. When tested with gammatone-filtered speech, 
single-channel experiments at a fixed spike budget demonstrate that the adaptive loop enhances onset responses while preserving fine temporal details 
that are lost in a fixed-\(\Delta\) baseline approach; multi-channel spike spectrograms confirm this advantage across frequency bands.

\smallskip
\noindent\textit{Limitations and future work —}
The current encoder IC and system are deliberately over-engineered to facilitate rapid development and proof-of-concept: 
they incorporate a discrete microcontroller, an external ArC TWO board, board-level analogue switches, and numerous support components, 
all of which introduce parasitic elements and compromise area and power efficiency. In the encoder IC, the on-chip threshold generator was intentionally designed with extended ranges—both for readable resistance and \(V_{\mathrm{read}}\)—to accommodate various memristor technologies. This versatility necessarily increases static read power consumption and silicon area requirements. 
Our future work aims to develop a low-power, highly scalable \emph{single-chip} hybrid CMOS–memristor implementation that co-integrates the memristor driver, switch matrix, and a lightweight on-chip controller. Where fabrication technology permits, we plan to monolithically integrate the memristor array on the same die\cite{maheshwari2021design}, which would enable more precise control while significantly reducing parasitics, area requirements, and power consumption.
This integrated approach would facilitate more sophisticated per-channel control (with programmable pulse parameters and adaptive comparison windows), higher channel counts to support demanding application scenarios,
minimise parasitic effects, and substantially improve both area efficiency and energy consumption per encoded spike, all while simplifying the overall system architecture.